\documentclass[preprint,amsmath,amssymb,aps,showkeys]{revtex4}

\usepackage{graphicx}       
\usepackage{bm}             
\usepackage{xcolor}         
\usepackage{mathtools}      
\usepackage{siunitx}        
\usepackage{multirow}       

\begin{document}
\title{Bistability in filamentous actin through monomer-sequestration of an effector species}

\author{Panayiotis Foteinopoulos}
\author{Bela M. Mulder}
\email{mulder@amolf.nl}
\affiliation{Institute AMOLF, Science Park 104, 1098XG Amsterdam, the Netherlands}%
\date{\today}

\begin{abstract}
Filamentous actin, a species of dynamic protein polymers, is one of the main components of the cytoskeleton of eukaryotic cells. We formulate a class of models that predict the possibility of bistable steady states in populations of dynamic actin filaments. They are built upon a basic model of actin dynamics that includes severing and capping in the presence of a finite actin monomer pool. The key additional ingredient is the presence of a single species of effector molecules that is partially sequestered to an inactive state by binding to free G-actin. In its unbound active state, this effector species can \emph{enhance} the rate of nucleation of filamentous actin or its growth speed, or \emph{inhibit} the activity of capping or severing proteins. Using an explicit analytical solution of the basic actin dynamics model, we show that bistability is predicted to occur in all of the proposed models. We verify these predictions using particle-based stochastic simulations. In addition, we show that switching between the two stable states can be achieved by transient manipulation of the free G-actin pool size.
\end{abstract}


\maketitle


\section{Introduction}\label{sec:intro}
Filamentous actin (F-actin) is one of the key components of the cytoskeleton of eukaryotic cells. Its functionality derives in part from its ability to form spatially extended dynamical structures, involved in maintaining the mechanical integrity of cells, mediating shape changes, and driving cell motility \cite{Blanchoin2014,Pollard2000}.  This versatility derives from the dynamic nature of the individual F-actin filaments, which can undergo continuous processes of nucleation, elongation, and shortening. Driving these (dis)assembly processes is the (de)association of the monomeric G-actin protein to the filaments.  The ATP-bound form of G-actin preferentially binds at the so-called barbed end of the F-actin filament. Within the filament, the ATP bound to the G-actin is hydrolyzed. At the other end of the filament, the so-called pointed end, the G-actins are then preferentially released, allowing the filaments as a whole to display treadmilling motion. 

In vivo, the dynamics of F-actin is regulated by a host of other proteins that influence assembly and disassembly mechanisms \cite{DosRemedios2003ActinMicrofilaments}, allowing cells to maintain different steady-state actin structures, but also undergo local and global cytoskeleton reorganization processes \cite{Blanchoin2014,Wales2016}. The monomer-binding protein profilin regulates the activity of competing nucleating proteins, such as the Arp2/3 complex and formins, and can enhance the polymerization speed at the growing end \cite{Suarez2015ProfilinComplex,Skau2015SpecificationFactors}.  On the other hand, capping proteins can inhibit G-actin polymerization by tightly binding to the barbed end \cite{Pollard2000}. Finally, proteins such as the ADF/Cofilin family and Gelsolin are central to ongoing actin dynamics and turnover control \cite{Carlsson2010ActinMicroscale,Ono2007MechanismDynamics}. These proteins assist in the recycling of actin back to the monomeric state by severing the filaments at some point along their length. 

Naturally, the rate of elongation, determined by the balance between barbed end growth and pointed end shrinkage, and the nucleation rate of F-actin filaments depend on the availability of ATP-bound G-actin in the cytoplasm. Biochemical studies revealed, for example, that the elongation rate depends linearly on the G-actin concentration \cite{Pollard1986RateFilaments}. Furthermore, the relative number of monomers tunes the distribution of actin in different F-actin structures \cite{Suarez2015ProfilinComplex,Burke2014HomeostaticMonomersb}. This means that in principle the dynamics of the actin cytoskeleton is also globally regulated by the necessarily finite total amount of G-actin in the cell. 

Recently, a number of studies have appeared that provide evidence that G-actin pool size limitations may play a prominent role in cellular processes. Lomakin et al.\ showed that competition for G-actin monomers between a population of bundled cortical actin and a rapidly growing dendritic network in IAR-2 epithelial cells allows switching between a static and a polarized and migratory cell morphology \cite{Lomakin2015}. More recent work has also highlighted the global role of monomer availability in regulating actin organization \cite{Suarez2016InternetworkOrganization,Carlier2017GlobalNetworks}. In mammalian culture cells, rapid breakdown of the cortical actin network coupled with rapid growth of an ER-associated perinuclear F-actin structure, followed by equally rapid recovery, can be observed after (bio)chemical and mechanical stimulation (first reported in \cite{Shao2015} in NIH 3T3 fibroblasts and later studied in depth on a large panel of cells in \cite{Wales2016}). Here again, the shift between the actin populations involved is mediated by the limited free G-actin pool.

The examples above involve two states that are distinguished by \emph{extrinsic} factors influencing the location and morphology of the actin networks involved. These two populations then compete for a limited pool of G-actin monomers and thus exert a mutually inhibitory effect on each other. Here we ask the question whether the pool size dependence of actin dynamics can be exploited to ``engineer'' \emph{intrinsic} bistability in an actin population, which only involves the actin dynamics itself.  To that end, we consider a class of models in which we introduce a single additional molecular agent that can modulate the underlying actin dynamics. This additional molecular species, which we dub an effector, can bind to the G-actin monomers, leading to its pool size-dependent sequestration. When coupled to the dependence on the G-actin concentration of both the growth and nucleation rates of F-actin, this mechanism effectively leads to the type of double inhibitory or activatory feedback motif that is well known to display bistability (for an overview, see \cite{Ferrell2002Self-perpetuatingBistability}). To implement these models, we build on a classical mesoscopic model for actin dynamics in the presence of severing \cite{Edelstein-Keshet1998,Ermentrout1998}. As it is more appropriate to this mesocopic population-level description, we extend this model to include the finite pool size effects in a phenomenological way, rather than considering more biochemically detailed \cite{Bindschadler2004} or stochastic \cite{Harbage2016} single-filament effects considered previously. Finally, we add the additional effector species, modeling both its dose-dependent effects on the actin dynamics and its binding equilibrium to the free G-actin pool.

\section{Model}
\subsection{Basic actin dynamics model}\label{sec:basic_model}
\subsubsection{Description}

We fix the total amount of actin in the system and express it as the aggregate length $L$ of all F-actin filaments when all available actin is fully polymerized. Similarly, the available G-actin monomer pool at every instant is expressed as the total length $L_{G}\left(t\right)$ of filaments that this pool would create if fully polymerized. Thus, actin mass conservation is described by the rule $L=L_F(t)+L_G(t)$, where $L_F(t)$ is the total filament length. The four components of actin dynamics included in the model are \emph{nucleation}, \emph{growth}, \emph{capping} and \emph{severing}.

New F-actin filaments are nucleated at a pool-size-dependent rate%
\begin{equation}
r_{n}\left(L_{G}\right)  =r_{n}^{\infty}\frac{L_{G}^{q}}{L_{G}^{q}+L_{\ast}^{q}},
\end{equation}
where $r_{n}^{\infty}$ is the maximal nucleation rate, $L_{\ast}$ a cross-over length and $q$ a Hill-coefficient that characterizes the degree of cooperativity of the nucleation process. The sigmoidal shape of the base dependence on the size of the pool models a strong dependence for $L_{G}/L_{\ast}\ll 1,$ where diffusion of monomers is the limiting factor, while allowing for saturation of the rate when $L_{G}/L_{\ast}\gg 1$, where the process is limited by an intrinsic rate. A priori, one expects the Hill coefficient for the nucleation process, which involves multiple monomers coming together, to be greater than unity. Here, however, since our primary focus is on distinguishing the behavior in the regime of pool size limitation to that in the saturation regime, and in the absence of relevant experimental data, we set $q=1$ throughout.

A newly nucleated filament grows in length at a speed $v_{+}\left(L_{G}\right)  =v_{b}\left(L_{G}\right) -v_{p}$, where $v_{b}$ is the pool size-dependent polymerization speed at the barbed end and $v_{p}$ the depolymerization speed at the pointed end. For a linear growth process, it is reasonable to assume the Hill coefficient of the dependence on the pool size to be unity, so we take the barbed-end polymerization speed to be
\begin{equation}
v_{b}\left(L_{G}\right)  =v_{b}^{\infty}\frac{L_{G}}{L_{G}+L_{\ast}},
\end{equation}
where $v_{b}^{\infty}$ is the maximum polymerization speed. Note that below a minimum size of the G-actin pool $L_{min} = \frac{v_p}{v_b^{\infty}-v_p}L_*$ the net growth speed is negative, and filaments cannot grow, as they disassemble faster than they can assemble.

The growing filaments can be capped at a rate $r_{c}$. After capping, the polymerization at the barbed end is suppressed and the filaments shrink in length with velocity $v_{-}=v_{p}$, ultimately releasing their entire length into the G-actin pool. 

Finally, filaments can be severed homogeneously along their length at a rate $r_{s}$ per unit length. We assume, with the known activity of gelsolin \cite{Sun1999GelsolinProtein} as an example, that the new barbed end created by the severing is immediately capped, so that the lagging strand is always in the shrinking state. The dynamical state of the leading strand is simply determined by the state of the original filament before the severing event, that is, growing if it was uncapped and shrinking if it was capped.

The dependent variables in our model are $a_{+}\left(  l,t\right)$ and $a_{-}\left(  l,t\right)$, the length distribution of the growing, respectively, shrinking, filaments at time $t$. Note that these are the unnormalized distributions that also carry the information on how many filaments there are and what the total length of filaments is. As the whole process conserves actin, the size of the monomer pool $L_{G}$ can be deduced when the latter two distributions are determined.

\subsubsection{Dimensional analysis and aggregate variables}
We adopt $l_0=L_*$ as our unit of length and $t_0=L_*/v_p$ as our unit of time, which sets the dimensionless length $\lambda=l/L_*$ and time $\tau=t_0 v_p/L*$. The dimensionless actin distributions are given by $\alpha_{\pm}(\lambda,\tau)=L_* a_{\pm}(l,t)$. Using these definitions we obtain the dimensionless nucleation rate and growth speed
\begin{align}
\nu\left(\Lambda_{G}\right)   &= \nu^{\infty} \frac{\Lambda_{G}}{\Lambda_{G}+1},\\
\omega_{+}\left(\Lambda_{G}\right) &= \omega^{\infty}\frac{\Lambda_{G}}{\Lambda_{G}+1}-1,
\end{align}
where $\Lambda_{G}=L_{G}/L_{\ast},$ $\nu^{\infty} =r_{n}^{\infty} L_{\ast}/v_p$ and $\omega^{\infty}=v_{b}^{\infty}/v_{p}$. 
Likewise, the dimensionless forms of the capping and severing rates are $\kappa= r_c L_*/v_p$ and $\sigma =r_s L^{2}_{\ast}/v_p$.

As aggregate variables, we first define the moments of the distributions through%
\begin{equation}
A^{(n)}_{\pm}\left(  \tau\right)  =\int_{0}^{\infty}d\lambda\,\lambda^{n}%
\alpha_{\pm}\left(  \lambda,\tau\right).
\end{equation}
The total number of F-actin polymers in a given state is given by $A_{\pm}(\tau)\equiv A^{(0)}_{\pm}(\tau)$ and the total length $\Lambda_{\pm}(\tau)\equiv A^{(1)}_{\pm}(\tau)$. Denoting the total polymerized length by $\Lambda_F(\tau)=\Lambda_{+}(\tau)+\Lambda_{-}(\tau)$, the conservation of actin length is expressed through $\Lambda=\Lambda_{G}(\tau)+\Lambda_{F}(\tau)$.

\subsubsection{Self-consistent solution}
The dynamical equations for actin in the presence of capping and severing in the case of an unlimited pool of G-actin were first formulated, but not explicitly solved, by Edelstein-Keshet and Ermentrout \cite{Edelstein-Keshet1998,Ermentrout1998}. These equations, in fact, also describe the dynamics of microtubules in the absence of so-called rescues, which mark the spontaneous switch from a shrinking state to a growing state in the microtubule dynamical instability mechanisms. In the latter context, an analytical solution to these equations was first obtained \cite{Tindemans2010a}. Here, we need to generalize these results to the current setting by including the dynamics of the monomer pool and its influence on the nucleation rate and growth speed. We relegate the technical details of the resultant derivation to the Appendix \ref{app:base_model}. The key result is that, in the steady state, the free G-actin length $\Lambda_G$ must satisfy a self-consistency relation imposed by the conservation of the total actin length $\Lambda$. The form of this self-consistency relation is most conveniently presented as
\begin{equation}\label{eq:self_consistency}
    \Lambda-\Lambda_G = \Lambda_F(\Lambda_G)=\nu(\Lambda_G)\Phi(\Lambda_G).
\end{equation}
The function $\Phi$, which heuristically can be interpreted as F-actin length contributed by each nucleation event, is explicitly given by
\begin{equation}
   \Phi(\Lambda_G)= \frac{1}{2\sigma}\sqrt{\pi} \frac{e^{\Omega^{2}\left(\Lambda_{G}\right)}%
\operatorname{erfc}\left(\Omega\left(\Lambda_{G}\right)\right)}{\Omega\left(\Lambda_{G}\right)},
\end{equation}
where
\begin{equation}
   \Omega\left(\Lambda_{G}\right) =\frac{\kappa}{\sqrt{2\sigma\omega_{+}\left(
\Lambda_{G}\right)  (\omega_{+}\left(  \Lambda_{G}\right)+1)}}. 
\end{equation}

\subsection{G-actin pool-dependent feedback mechanisms}
\subsubsection{Effector species sequestered by G-actin}
We now introduce an additional species present in a fixed and large total number $B$, allowing us to treat it as a continuous variable. This species can be sequestered by binding to actin monomers, resulting in an \emph{inactive state} $0$. When unbound, it is in an \emph{active state} $1$. We assume that this species is fast-diffusing, so that we can assume well-mixing, allowing us to use simple global binding dynamics
\begin{align}
    \frac{d}{dt} B_1(t) &= k_u B_0(t)-k_b B_1(t) L_G(t), \\
    \frac{d}{dt} B_0(t) &= -k_u B_0(t)+k_b B_1(t) L_G(t),
\end{align}
where $B_1$ is the number of active, unbound, effectors, $B_0$ the number of inactive, bound, effectors, $k_b$ is the rate of binding of the effector to G-actin per unit of length and $k_u$ the rate of unbinding.  In steady state we have the simple binding equilibrium
 \begin{equation}
     B_1 = \frac{k_u}{k_u+k_b L_G} B.
 \end{equation}
Introducing the fraction of effectors in the active state through $\beta = B_1/B$, we can write
 \begin{equation}\label{eq:freeB}
     \beta(\Lambda_G) = \frac{\Lambda_{d}}{\Lambda_{d}+\Lambda_G}
 \end{equation}
where $\Lambda_{d}=k_u/(k_b L_*)$ is interpreted as the dimensionless dissociation constant governing the affinity of the species $B$ to the G-actin monomers.

\subsubsection{Feedback mechanisms}
In its active state, the effector modulates a target parameter which we will choose to be one of the dynamical parameters $\{r_n^{\infty},v_b^{\infty},r_c, r_s\}$. Denoting the modulated parameter by $x$, the degree of modulation follows a generic non-linear dose-response curve: 
\begin{equation}
    x(\beta) = \frac{\beta_*^h(1-\beta^h)}{\beta_*^h(1-\beta^h)+\beta^h(1-\beta_*^h)} x_0+\frac{\beta^h(1-\beta_*^h)}{\beta_*^h(1-\beta^h)+\beta^h(1-\beta_*^h)} x_1, \label{eq:modulation}
\end{equation}
 where $\beta_*$ sets the scale of the Hill-type dose-response curve, and $h$ is a Hill parameter that controls the steepness of the cross-over between the low effect ($0 \le  \beta < \beta_*$) and the high effect ($\beta_* < \beta \le 1$) regime. Note that this form implicitly assumes that there are a sufficient number of effectors and/or their activity is high enough such that they are effective in modulating the actin dynamics.  This choice is pragmatic and obviates the need for a separate analysis of the dependence on the absolute number of effectors.   
 
 Specifically, the four types of modulation we consider are the following:
 \paragraph{Nucleation rate} In this case, the base maximal nucleation rate $r_n^{\infty}$ is modulated so that $r_n^{\infty}(0) < r_n^{\infty}(1)$. Therefore, the presence of an active effector enhances the production of novel F-actin filaments, leading to more filaments. 
\paragraph{Polymerization speed}  In this case, the base polymerization speed at the barbed end $v_b^{\infty}$ is modulated so that $v_b^{\infty}(0) < v_b^{\infty}(1)$. Therefore, the presence of an active effector enhances the growth of F-actin, leading to longer filaments. 
\paragraph{Capping rate} In this case, the capping rate $r_c$ is modulated so that $r_c(0) > r_c(1)$. Therefore, the presence of an active effector suppresses the capping of F-actin filaments, leading to longer filaments. 
\paragraph{Severing rate} In this case, the capping rate $r_s$ is modulated so that $r_s(0) > r_s(1)$. Therefore, the presence of an active effector suppresses the severing of F-actin filaments, leading to longer filaments. 

\begin{figure}
    \centering
    \includegraphics[width=\textwidth]{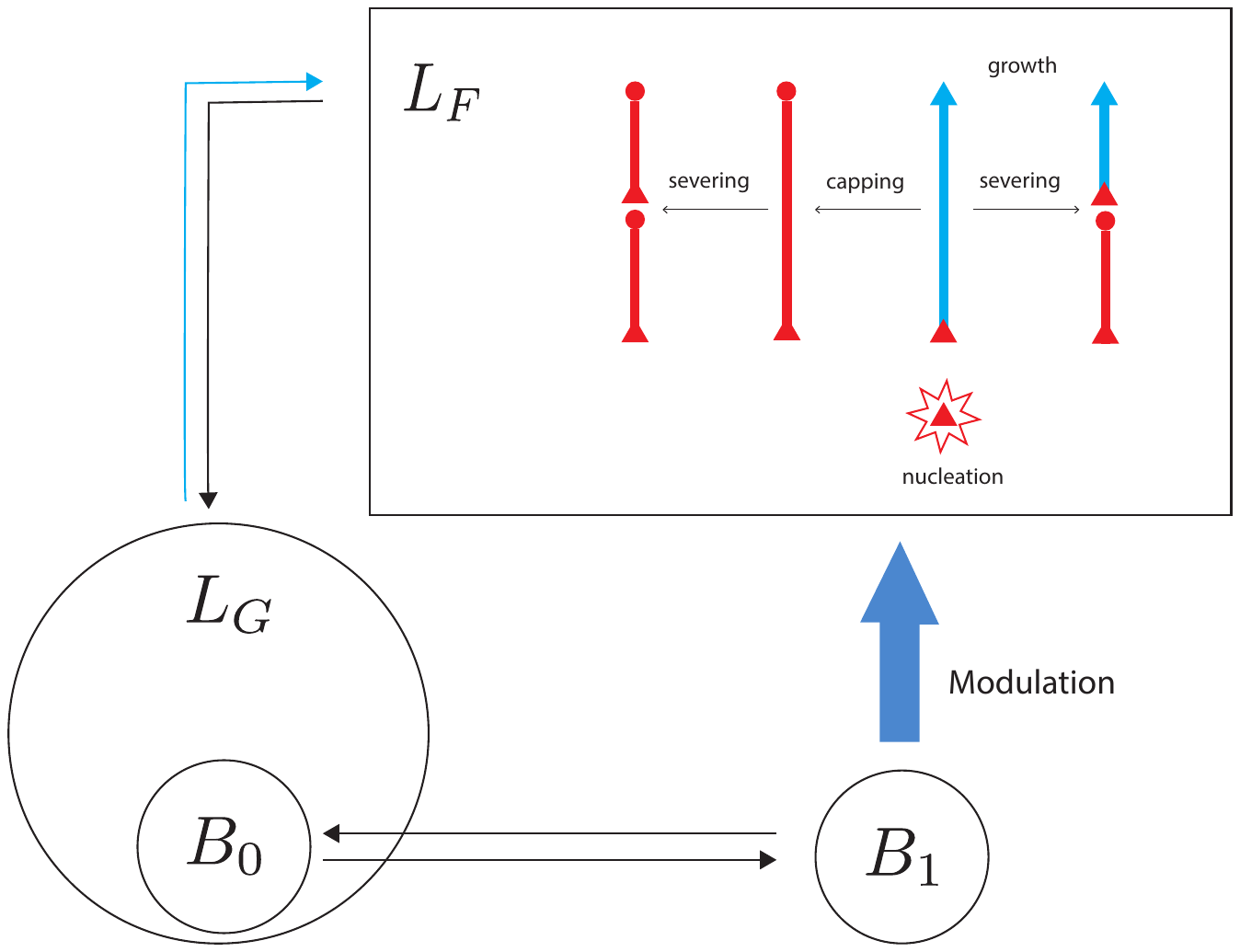}
    \caption{Schematic of our actin dynamics model. The polymerized F-actin filaments with total length $L_F$ exchange monomers with the free G-actin pool of size $L_G$, which by itself controls both the barbed end polymerization speed of growing filaments (cyan) and the nucleation rate of new filaments. The free actin pool also sequesters the molecular effector species $B$ into an inactive fraction $B_0$. The remaining active fraction $B_1$ can mediate actin dynamics either by enhancing the nucleation rate or the growth speed, or by decreasing the capping or the severing rate. }
    \label{fig:schematic}
\end{figure}

In all of these four cases, we expect that a large fraction of active effector is associated with a large fraction of polymerized F-actin, and hence a small fraction of free G-actin, which in turn is consistent with a small fraction of bound inactive effector. On the contrary, a small fraction of active effector is associated with a small fraction of polymerized F-actin, and hence with a large fraction of free G-actin, again consistent with a large fraction of bound inactive effector. This has all the hallmarks of a generic mutual-repression motif (e.g.,see \cite{Chen2012Sequestration-basedLatch}), well known to lead to bistability. The common structure of these models is schematically illustrated in Figure \ref{fig:schematic}.

\subsection{Overview of model parameters}
Having set up our model above, we provide an overview of the parameters in Table \ref{tab:parameters}.
\begin{table}[h]
    \centering
    \begin{tabular}{cc>{\centering}p{3cm}l}
    \hline
        Type & Parameter & Definition & Description\\
        \hline
         {\multirow[b]{2}{1cm}{\rotatebox{90}{Units}}} & $L_*$   & & Cross-over length pool-size dependence basic actin dynamics \\
                 & $v_p$ & & Depolymerization speed\\
         \hline
        {\multirow[c]{5}{1cm}{\rotatebox{90}{{Baseline model}}}} & $\nu^\infty$          & $r_n^\infty L_*/v_p$      & Bare nucleation rate \\
                        & $\omega^\infty$       & $v_b^\infty/v_p$          & Bare poymerization speed \\
                        & $\kappa$              & $r_c L_*/v_p$             & Capping rate \\
                        & $\sigma$              & $r_s L_*^2/v_p$           & Severing rate \\
                        & $\Lambda$             & $L/L_*$                   & Total actin length \\

        \hline
        {\multirow [c]{7}{1cm}{\rotatebox{90}{Modulation}}} & $\Lambda_{d}$ & $k_u/(v_p L_*)$ & Dissociation constant of effectors\\
                    & $\beta_{*}$ &                  & Cross-over value of the effector dose-response curve \\
                    & $h$         &                  & Hill coefficient effector dose-response curve \\
                    & $\nu^\infty(0),\nu^\infty(1)$   &             & min,max bare nucleation rate\\
                    & $\omega^\infty(0),\omega^\infty(1)$   &       & min,max bare polymerization speed\\
                    & $\kappa(0),\kappa(1)$   &                     & max,min capping rate\\
                    & $\sigma(0),\sigma(1)$   &                     & max,min severing rate\\    
        \hline
    \end{tabular}
    \caption{Overview of the parameters of our model}
    \label{tab:parameters}
\end{table}

\section{Results}\label{sec:results}
\subsection{Minimal scenario: Modulation of nucleation in the absence of severing}\label{sec:results_minimal}
\subsubsection{Analytical analysis}
First, we explore a minimal scenario in which the effector modulates the bare nucleation rate for actin dynamics in the absence of severing. Moreover, we assume that the modulation is hypersensitive to the fraction of free effectors $\beta$, that is, the Hill coefficient in Eq.\ (\ref{eq:modulation}) that governs the dose-response curve $h \rightarrow \infty$.

Under the hypersensitivity assumption the bare nucleation rate behaves as
\begin{equation}
    \nu^\infty(\beta) = \nu_0 \theta(\beta_*-\beta)+\nu_1 \theta(\beta-\beta_*),
\end{equation}
where $\theta(.)$ is the Heavyside step function, and $\nu_1 > \nu_0$. Given the dependence of the free effector fraction on the size of the G-actin pool Eq.\ (\ref{eq:freeB}), we can eliminate $\beta$ to obtain the full dependence of the nucleation rate on the G-actin pool size 
\begin{equation}
    \nu(\Lambda_G) =\left( \nu_1 \theta(\Lambda_*-\Lambda_G)+\nu_0 \theta(\Lambda_G-\Lambda_*)\right) \frac{\Lambda_G}{\Lambda_G+1},
\end{equation}
where the critical pool size is given by $\Lambda_* = \frac{1-\beta_*}{\beta_*}\Lambda_d$. We now note that for any total amount of actin $\Lambda$, a dissociation constant $\Lambda_d$ may be chosen such that $\Lambda_* < \Lambda$.

We then write the self-consistency equation Eq.\ (\ref{eq:LamSS}) in the no-severing limit Eq.\ (\ref{eq:Psi0}) as
\begin{align}\label{eq:ultra}
\Lambda-\Lambda_G  &= \nu(\Lambda_G)\Phi_0(\Lambda_G) \nonumber \\
                   &= \frac{\omega^\infty}{\kappa^2} \left( \nu_1 \theta(\Lambda_*-\Lambda_G)+\nu_0 \theta(\Lambda_G-\Lambda_*)\right) \left(\frac{\Lambda_G}{\Lambda_G+1}\right)^2 \left( \omega^\infty \frac{\Lambda_G}{\Lambda_G+1}-1\right) \nonumber \\
                   &\equiv \left( \hat{\nu}_1 \theta(\Lambda_*-\Lambda_G)+\hat{\nu}_0 \theta(\Lambda_G-\Lambda_*)\right)\phi_0(\Lambda_G),
\end{align}
where we have absorbed all multiplicative constants into the rescaled nucleation rates $\hat{\nu}=\omega^\infty/\kappa^2 \nu$. The function $\phi_0$ has the following relevant characteristics: (i) $\phi_0(\Lambda_{min})=0$, where $\Lambda_{min}=\frac{1}{\omega^\infty-1}$ is the minimal size of the G-actin pool to sustain the growth of filaments, (ii) $\phi_0'(\Lambda_G) >0$ for $\Lambda_G >\Lambda_{min}$, that is, it is positive and monotonically increasing. For the model to make sense, we need $\Lambda > \Lambda_* > \Lambda_{min}$.

\begin{figure}
    \centering
    \includegraphics[width=\textwidth]{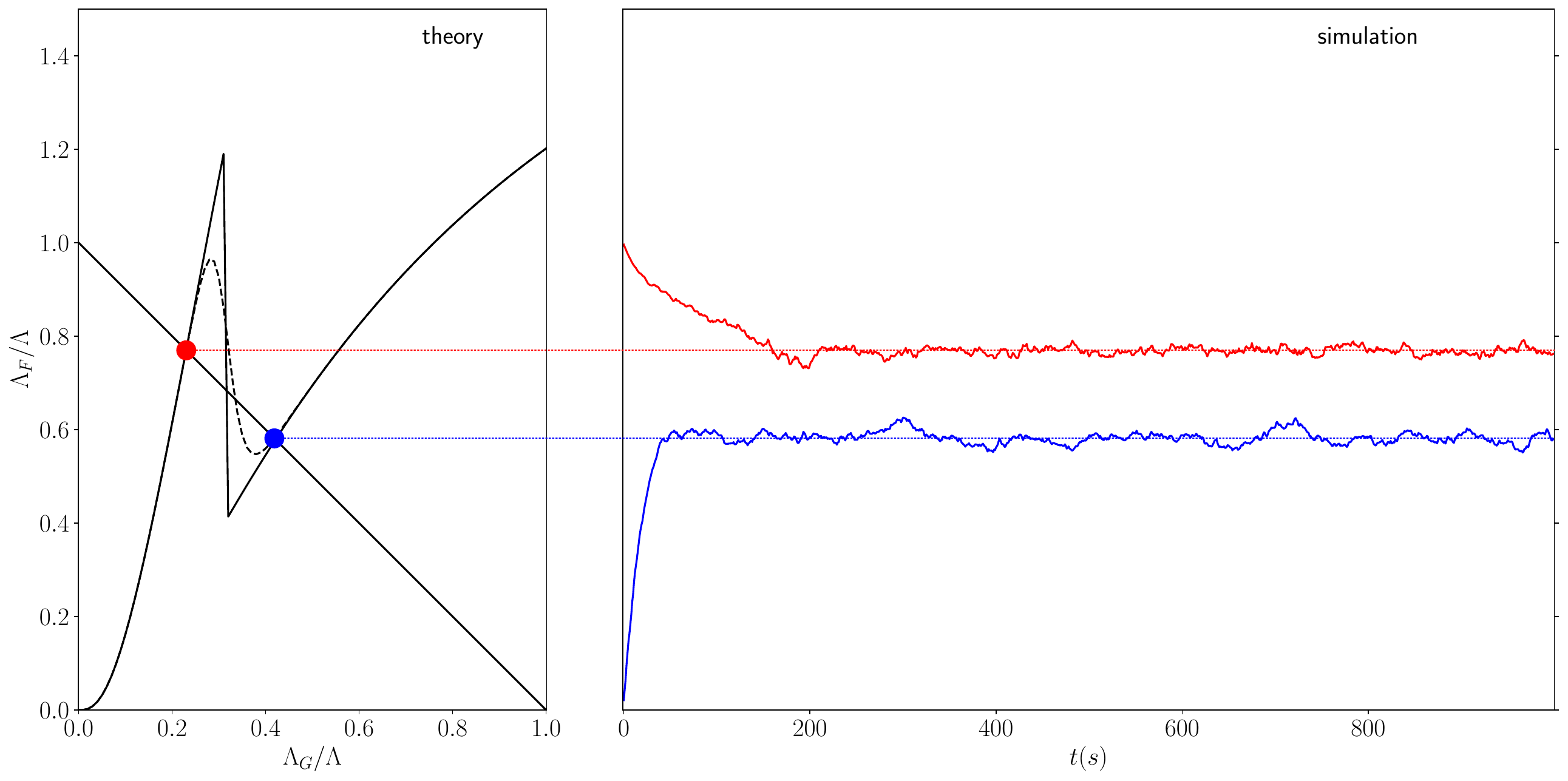}
    \caption{Left-hand panel: Graphical representation of the self-consistency condition in the minimal scenario, showing the two stable states. Solid line: the ultrasensitive case $h \rightarrow \infty$ (LHS Eq.\ (\ref{eq:ultra})), dashed line: finite non-linearity case $h = 20$ (LHS Eq.\ (\ref{eq:self_consistency})). Right-hand panel: Results of simulations starting from either a fully polymerized state (red curve), or a fully depolymerized state (blue curve).}
    \label{fig:minimal}
\end{figure}

If we now choose $\hat{\nu}_1> \hat{\nu}_0$ so that $\hat{\nu}_1 \psi_0(\Lambda_*) > \Lambda-\Lambda_*$ and $\hat{\nu}_0 \psi_0(\Lambda_*) < \Lambda-\Lambda_*$, then the graphical representation of the self-consistency condition has the form shown in the left-hand panel of Figure \ref{fig:minimal}. This clearly shows that in this case there are two stable states, the first with a higher value of the polymerized length, the second with a lower one. The two parameters that basically determine whether the bistability occurs or not are (i) the dissociation constant $\Lambda_d$, which determines whether the (un)binding of the effectors from the monomers is salient for the total amount of actin and (ii) the ratio of the effective nucleation rates $\hat{\nu}_1/\hat{\nu}_0$, which opens the `gap' between the stable states. We also show that one can relax the requirements on the Hill coefficient $h$: for a value of $h=20$ the predicted values of the bistable states differ only marginally from those of $h\rightarrow\infty$. 

Generally, once the right-hand side of the self-consistency equation Eq.\ (\ref{eq:self_consistency}) has an inflection point with negative derivative, bistability can be achieved. We can then either raise or lower the nucleation rate by a common factor, moving the curve up or down to ensure intersection with the linear left-hand side. Alternatively, we can raise or lower the total amount of actin $L$, or equivalently lower or raise the cross-over length $L_*$ to achieve the same effect by raising or lowering the left-hand side to ensure intersection with the right-hand side. 

\subsubsection{Stochastic simulations}
In order to validate our analytical predictions, we also performed stochastic simulations of our dynamical actin model. In these finite-time step simulations the system evolves from an initial state to state, which we can compare with the predicted stable states. Starting either from a fully polymerized state or a fully depolymerized initial state, we can select whether the system evolves towards the state with the higher- or the lower total amount of F-actin respectively. For details on the implementation of these simulations and the parameters employed, we refer the reader to Appendix \ref{app:simulations}. The results, presented in the right-hand panel of Fig.\ \ref{fig:minimal}, show that quantitative agreement with the predictions is obtained. We also tested whether we can further decrease the non-linearity of the dose-response curve of the effect of the effector binding on the nucleation rate. We were able to achieve bistability for a Hill coefficient of $h=8$ keeping all dynamical parameters at their baseline values, except for the total amount of actin, which had to be decreased by a factor close to two (data not shown). 
\subsubsection{Switching between states induced by transient pool manipulation}
Having shown that bistability readily occurs in our model, we ask whether it is possible to switch between the two stable states by manipulating the total amount of actin available. The logic is as follows: if we increase the total amount of free G-actin, the binding equilibrium of the effector molecules will be shifted towards the bound inactive state, which will destabilize the more highly polymerized state. Conversely, if we decrease the total amount of free G-actin, we release effector molecules into the active state, hence destabilizing the state with the lower degree of polymerization. By applying these changes only transiently for a duration $\Delta \tau$, the system is restored to its original total amount of actin so that the original pair of stable states is again salient. In Figure \ref{fig:simple_switching}, we show the results of a series of such simulations (see Appendix \ref{app:simulations} for details).
\begin{figure}
    \centering
    \includegraphics[width=\textwidth]{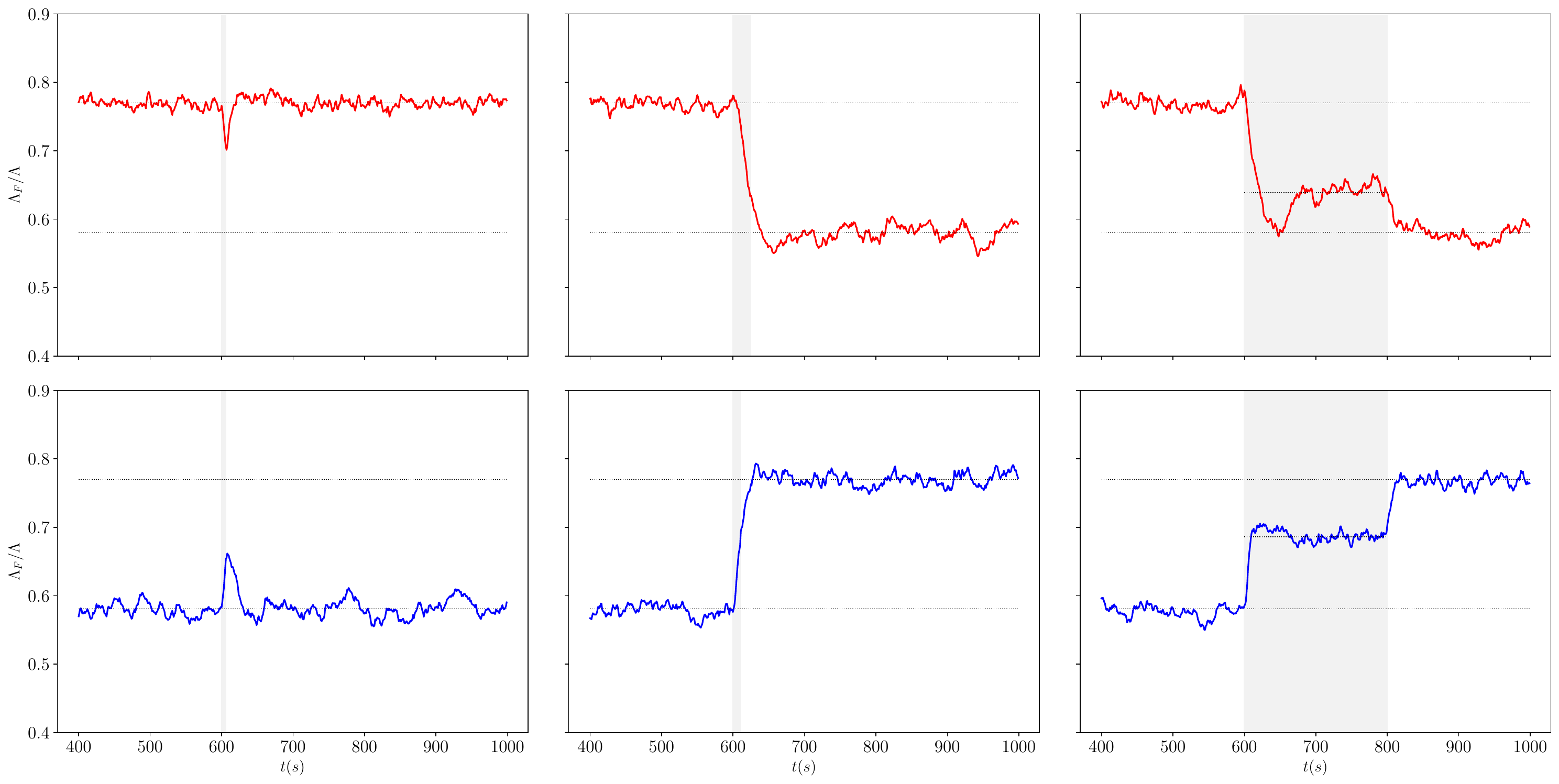}
    \caption{Switching between bistable states by transiently manipulating the available G-actin pool. Top row: $\delta \Lambda_{\downarrow}=$\SI{10}{\percent}, $\Delta\tau_{\downarrow}$ from left to right: \SI{6}{\second}, \SI{25}{\second}, \SI{200}{\second}. Bottom row: $\delta \Lambda_{\uparrow}=$\SI{-10}{\percent}, $\Delta\tau_{\uparrow}$ from left to right: \SI{6}{\second}, \SI{11}{\second}, \SI{200}{\second}. Dotted lines: predicted stable/intermediately stable states.}
    \label{fig:simple_switching}
\end{figure}
We see that if the duration $\Delta\tau$ of the manipulation of the actin availability is shorter than a characteristic reaction time of the system, switching cannot occur. For intermediate $\Delta\tau$, the system switches essentially monotonically. Finally, for larger $\Delta\tau$, the system can first reach an intermediate state adapted to the changed total amount of actin, which subsequently decays to the new final state once the manipulation stops. Note also that switching from the low-polymerizated state to the high-polymerized state is faster by an order of magnitude than the other way around. This reflects the fact that the maximal barbed-end polymerization speed, which drives the build-up to the high polymerized state, is much larger than the pointed-end shrinking speed of actin, which drives the breakdown of the high polymerized state.   
\subsection{The four scenarios, including severing}\label{sec:results_full}
\begin{figure}
    \centering
    \includegraphics{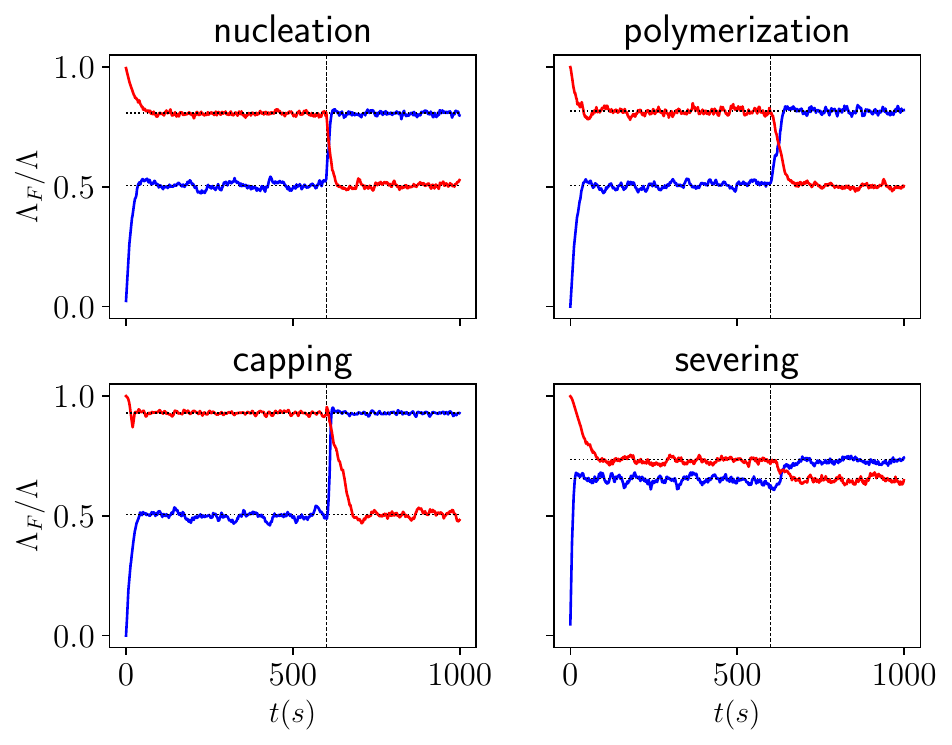}
    \caption{Results showing bistability and state-switching induced by transient monomer pool manipulation for the full model including severing. Dotted lines are the predicted locations of the stable states. The values for the modulated parameters used are tabulated in Table \ref{tab:modulated_pars}. The dashed line at $t=$\SI{600}{\second} indicates the start of the G-actin pool size manipulation. The duration and magnitudes of these manipulations are collected in Table \ref{tab:pool_manipulation} in Appendix \ref{app:simulations}.}
    \label{fig:full_modulation}
\end{figure}

In Fig.\ \ref{fig:full_modulation} we show both bistability and the ability to switch between the bistable states using transient manipulation of the monomer density for all four scenarios in the full model including severing: modulating the nucleation rate, the barbed-end polymerization rate, the capping rate and the severing rate, respectively. 

The limiting values for the modulated parameters are given in Table \ref{tab:modulated_pars}.
\begin{table}   
\begin{tabular}{llll}
\hline\noalign{\smallskip}
Parameter & $x(\beta)$ &  $x(0)$ & $x(1)$\\
\noalign{\smallskip}\hline\noalign{\smallskip}
Bare nucleation rate          & $r_{n}^\infty(\beta)$   & \SI{70}{\per\second}   & \SI{350}{\per\second} \\
Bare polymerization speed     & $v_b^\infty(\beta)$     & \SI{15.6}{\micro\meter\per\second} & \SI{45}{\micro\meter\per\second} \\
Capping rate                  & $r_c(\beta)$            & \SI{3}{\per\second}  & \SI{0.3}{\per\second}\\
Severing rate                 & $r_s(\beta)$            & \SI{0.040}{\per\second\per\micro\meter} & \SI{0.005}{\per\second\per\micro\meter} \\
\noalign{\smallskip}\hline
\end{tabular}
\caption{\label{tab:modulated_pars} Values for the modulated parameters in the four scenarios leading to bistability.}
\end{table}
In all cases, we needed at most a modulation of one order of magnitude to achieve bistability. In all cases, except for the modulation of severing, the unmodulated value of the parameter in question could be taken equal to that of the reference model (see Table \ref{tab:baseline_pars}). For the case of severing, it appears that the baseline value of $r_s(0)=$\SI{0.005}{\per\second\per\micro\meter} is too low to meaningfully influence the state of the system when modulated downward. In this case, we therefore adopted an $8$-fold higher unmodulated value, modulating downward to the baseline value. Even then, however, the bistability gap is small compared to the other cases. 
\section{Discussion}
We have shown that \emph{in principle} a population of F-actin could exhibit bistability. To achieve bistability, we needed two ingredients. The first ingredient is the well-established biochemical fact that the dynamics of F-actin through the nucleation rate and the polymerization speed is explicitly dependent on the availability of G-actin monomers. The second ingredient is an effector species that has two key properties: (i) in its active state it influences the dynamics of F-actin and (ii) it is sequestered into an inactive state by binding to the G-actin monomers. Although there are a number of species that have the former property (for an overview, see \cite{Pollard2016ActinProteins}), the latter property appears limited to a few proteins that contain the so-called RPEL motif \cite{Mouilleron2008MolecularMAL, Wiezlak2012G-actinAssembly}. We are currently not aware of a species that combines both properties. This raises two questions. 

The first question is whether our proposed mechanism actually is realized \emph{in vivo}. In this context, we note that the current work was inspired by the Calcium-mediated Actin Reset (CaAR) mechanism \cite{Wales2016}. This is an adaptive mechanism by which a class of mammalian cells responds to external stress signals, which may be chemical or mechanical, by a temporary breakdown of their actin cortex. This breakdown is caused by the transient activity of a strong nucleator INF2, which is associated with the perinuclear ER. This effectively depletes the G-actin monomer pool that sustains the F-actin cortex, causing its breakdown. After a few minutes, these effects die out and the actin cortex reestablihes itself, but it is unclear whether its properties were in fact identical to those of its prestimulus state. This led us to the question of whether, in principle, bi-stability could occur in F-actin populations, with the CaAR mechanism tripping the switch between the two stable states.  At the same time, it was shown that one of the downstream effects of the CaAR response was mediated by a transcription factor that was released after being sequestered by being bound to G-actin monomers. This suggested to us the possible relevance of an effector being released by the transient depletion of the free G-actin population. However, to establish bistability of the type discussed in the current work, ideally one would need access to the length distribution of the F-actin \emph{in vivo}, as this is the primary distinguishing characteristic of stable states. This is experimentally challenging, and most results on this issue are obtained from \emph{ex vivo} work using cell extracts or reconstructed solutions \cite{Burlacu1992DistributionMicroscopy,Biron2004TheF-Actin}.  An alternate but slightly weaker reporter is the actin turnover time, equal to the total F-actin length divided by the number of shrinking F-actins, which equals the time it would take to fully depolymerize the current polymerized actin length. This quantity should be estimable using, e.g., turnover of an optogenetically activated F-actin-binding fluorophore.

The second question is to see whether it would be possible in the spirit of synthetic biology to engineer such a system \emph{ex vivo}. Here, there may be cause for cautious optimism. Firstly, the basic dynamical parameters that we used throughout (see Table \ref{tab:baseline_pars}) are either literature values or, when these were unavailable, reasonable estimates that produce feasible F-actin populations of order $10^2-10^3$ filaments. Secondly, engineering a chimeric construct that has the dual properties of binding both to G-actin and to F-actin and effecting some change in the dynamics of the latter may well be possible, as the relevant molecular biology techniques have been around for a while (for a review, see \cite{Yu2015SyntheticApplications}). Lastly, we observed that the degree of modulation of F-actin dynamical parameters necessary to achieve bistability was at most an order of magnitude, which may be feasible. The critical factor in this endeavor may be the relatively high degree of cooperativity of the effector-induced modulation of the F-actin dynamics ($\sim \mathcal{O}(10)$) that we found required. Here, a more systematic exploration of the parameter space than we opted for in this proof-of-principle study would be useful.

Finally, it is interesting to speculate whether the type of bistability described here could also occur in populations of microtubules whose dynamics is very similar to that of F-actin. If such a mechanism could be coupled to post-tranlational modifications, it could play a role in situations where distinct subpopulations of microtubules coexist, e.g., in neurites \cite{Iwanski2023CellularCytoskeleton} and mitotic spindles \cite{Tipton2022MoreSpindle}.

\begin{acknowledgments}
    We would like to thank Roland Wedlich-S\"{o}ldner (M\"{u}nster) for fruitful discussions and Daan Mulder (AMOLF) for a critical reading of the manuscript. We acknowledge preliminary work by MSc students Ireth Garc\'{i}a Aguilar (TU Delft) on dynamics with a finite G-actin pool, and Daniel Kloek (Refined, Malm\"{o}) and Tom van der Mijn (ToetsPers, Amsterdam) on bistability. The work of B.M.M. is part of the Dutch Research Council (NWO) and was performed at the research institute AMOLF.
\end{acknowledgments}
\appendix
\section{The generalized Edelstein-Keshet model and its solution}\label{app:base_model}

\subsection{Dynamical equations}

The dynamical equations for actin in the presence of capping and severing in the case of an unlimited pool of G-actin were first formulated by Edelstein-Keshet and Ermentrout \cite{Edelstein-Keshet1998,Ermentrout1998}. Here, we generalize these results to our setting by including the dynamics of the monomer pool and its influence on the nucleation rate and growth speed. 

To formulate the equations, we first define the complement of the cumulative distributions%
\begin{equation}
\hat{A}_{\pm}\left(\lambda,\tau\right)  =\int_{\lambda}^{\infty}d\lambda^{\prime}
\,\alpha_{\pm}\left(  \lambda^{\prime},\tau\right),
\end{equation}
from which the distributions themselves follow by differentiation%
\begin{equation}
\alpha_{\pm}\left(\lambda,\tau\right)  =-\frac{\partial}{\partial\lambda
}\hat{A}_{\pm}\left(\lambda,\tau\right).
\end{equation}
Note that
\begin{equation}
    \hat{A}_{\pm}(0,\tau) = A^{(0)}_{\pm}(\tau) = A_{\pm}(\tau)
\end{equation}
also defines the total number of growing or shrinking filaments.

With the definitions given above, the dynamical equations read
\begin{align}
\frac{\partial}{\partial\tau}\alpha_{+}(\lambda,\tau) & =-\omega_{+}\left(
\Lambda_{G}\left(\tau\right)\right)  \frac{\partial}{\partial\lambda
}\alpha_{+}(\lambda,\tau)-\kappa \alpha_{+}(\lambda,\tau) -\sigma \lambda\alpha_{+}(\lambda,\tau)
+\sigma \hat{A}_{+}\left(  \lambda,\tau\right)  \label{eq:alphap}%
\\
\frac{\partial}{\partial\tau}\alpha_{-}(\lambda,\tau)  &= \frac{\partial
}{\partial\lambda}\alpha_{-}(\lambda,\tau)+\kappa\alpha_{+}(\lambda,\tau
)-\sigma\lambda\alpha_{-}(\lambda,\tau)+2\sigma \hat{A}_{-}\left(  \lambda
,\tau\right)
  +\sigma \hat{A}_{+}\left(  \lambda,\tau\right) \label{eq:alpham}%
\end{align}
Thes e equations are supplemented by the nucleation boundary condition%
\begin{equation}
\omega_{+}\left(\Lambda_{G}\left(  \tau\right) \right)  \alpha_{+}%
(0,\tau)=\nu\left(  \Lambda_{G}\left(  \tau\right) \right).
\end{equation}

To understand the appearance of the complement to the cumulative length distributions as gain terms in the equations for the length densities, consider the rate at which, for example, a growing filament of length $\lambda$ is produced as the result of a severing event. This happens when a growing filament of length $\ \lambda^{\prime}>\lambda$ is severed, producing a growing leading strand. The rate at which such a filament is severed $\propto\sigma\lambda^{\prime}.$ The probability density that upon severing a leading strand of length $\lambda$ is produced $p\left(  \lambda|\lambda^{\prime}\right)  =\left(  \lambda^{\prime}\right)^{-1}$ due to the assumed uniformity of the severing. Thus, the total rate at which growing filaments are produced by severing events is%
\begin{equation}
J_{+}\left(  \lambda,t\right)  =\int_{\lambda}^{\infty}d\lambda^{\prime
}\,\frac{1}{\lambda^{\prime}}\times\sigma\lambda^{\prime}\times\alpha
_{+}(\lambda^{\prime},\tau) =\sigma\int_{\lambda}^{\infty}d\lambda^{\prime
}\alpha_{+}(\lambda^{\prime},\tau)=\sigma \hat{A}_{+}\left(\lambda,\tau\right).
\end{equation}

The dynamics of the G-actin pool is simply given by the balance between gain through depolymerization and loss through polymerization of F-actin
\begin{equation}\label{eq:LamG}
\frac{d}{d\tau}\Lambda_{G}\left(\tau\right) = A_{-}\left(  \tau\right)  -\omega_{+}\left(
\Lambda_{G}\left(\tau\right)\right)A_{+}\left(\tau\right), 
\end{equation}
Finally, the binding and unbinding of the effector species to the G-actin leads to
\begin{equation}
\frac{d}{d\tau}\beta(\tau) = \Lambda_d \left(1-\beta(\tau)\right) - \beta(\tau)\Lambda_G(\tau).
\end{equation}

\subsection{Moment equations}
We obtain the equations for the moments by multiplying Eqs.\ (\ref{eq:alphap}) and
(\ref{eq:alpham}) by $\lambda^{n}$ and integrating over $\lambda$. This yields%
\begin{eqnarray}
\frac{d}{d\tau}A^{(n)}_{+}\left(  \tau\right)   
&=&
\omega_{+}\left(
\Lambda_{G}\left(  \tau\right)  \right)  \delta_{n,0}\alpha_{+}(0,\tau
)+n \omega_{+}\left(\Lambda_{G}\left(\tau\right)\right) A^{(n-1)}%
_{+}\left(\tau\right)   \nonumber \\ 
&& -\kappa A^{(n)}_{+}\left(  \tau\right)  -\sigma A^{(n+1)}%
_{+}\left(\tau\right)  +\frac{1}{n+1}\sigma A^{(n+1)}_{+}\left(  \tau\right)
\\
\frac{d}{d\tau}A^{(n)}_{-}\left(\tau\right)   &=&
-\delta_{n,0}\alpha
_{-}(\lambda,\tau)-n A^{(n-1)}_{-}\left(\tau\right)  + \kappa A^{(n)}_{+}\left(
\tau\right) -\sigma A^{(n+1)}_{-}\left(  \tau\right)  
\nonumber \\
&&
+\frac{2}{n+1}\sigma
A^{(n+1)}_{-}\left(  \tau\right)  +\frac{1}{n+1}\sigma A^{(n+1)}_{+}\left(
\tau\right).
\end{eqnarray}
It is immediately apparent that, due to the presence of severing, the moments are coupled in the forward direction, so this system does not admit a closed solution based on a finite number of moments. 

We nevertheless consider the first two moment equations separately, as they are useful in the analysis of the steady-state solution below.

For $n=0$ we find%
\begin{align}
\frac{d}{d\tau}A^{(0)}_{+}\left(  \tau\right)   &  =\omega_{+}\left(
\Lambda_{G}\left(  \tau\right)  \right)  \alpha_{+}(0,\tau)-\kappa A^{(0)}_{+}\left(
\tau\right) \label{eq:M0p}\\
\frac{d}{d\tau}A^{(0)}_{-}\left(  \tau\right)   &  =-\alpha_{-}(\lambda
,\tau)+\kappa A^{(0)}_{+}\left(  \tau\right)  +\sigma\left\{  A^{(1)}_{+}\left(
\tau\right)  +A^{(1)}_{-}\left(  \tau\right)  \right\}. 
\label{eq:M0m}%
\end{align}
 
For $n=1$ we have%
\begin{align}
\frac{d}{d\tau}A^{(1)}_{+}\left(\tau\right)   &  =\omega_{+}\left(
\Lambda_{G}\left(\tau\right)\right)A^{(0)}_{+}\left(\tau\right)
-\kappa A^{(1)}_{+}\left(\tau\right) -\frac{1}{2}\sigma A^{(2)}_{+}\left(\tau\right)
\label{eq:M1p}\\
\frac{d}{d\tau}A^{(1)}_{-}\left(\tau\right)   &  =-A^{(0)}_{-}\left(
\tau\right)  +\kappa A^{(1)}_{+}\left(\tau\right)  +\frac{1}{2}\sigma A^{(2)}%
_{+}\left(  \tau\right).  \label{eq:M1m}%
\end{align}
The latter equations allow explicit verification of the conservation of total actin length, as (cf.\ Eq.\ (\ref{eq:LamG}))%
\begin{equation}
\begin{split}
\frac{d}{d\tau}\left\{  \Lambda_{+}\left(  \tau\right)  +\Lambda_{-}\left(
\tau\right)  \right\} &{}=\frac{d}{d\tau}A^{(1)}_{+}\left(  \tau\right)  +\frac
{d}{d\tau}A^{(1)}_{-}\left(  \tau\right) \\
&{}=\omega_{+}\left(  \Lambda_{G}\left(
\tau\right)  \right)  A^{(0)}_{+}\left(  \tau\right)  -A^{(0)}_{-}\left(
\tau\right)  =-\frac{d}{d\tau}\Lambda_{G}\left(  \tau\right).
\end{split}1
\end{equation}

\subsection{Steady state solution}

To study the steady-state solutions it suffices to note that in the steady state the pool size is fixed to an, as yet undetermined, value $\bar{\Lambda}_{G},$ where we will use the overbar for all quantities dependent on this value. In steady state the equations (\ref{eq:alphap}) and (\ref{eq:alpham}) become%
\begin{align}
\bar{\omega}_{+}\frac{d}{d\lambda}\alpha_{+}(\lambda)  &  =-\kappa \alpha_{+}%
(\lambda)-\sigma\lambda\alpha_{+}(\lambda)+\sigma \hat{A}_{+}\left(  \lambda\right)
\label{eq:apSS}\\
-\frac{d}{d\lambda}\alpha_{-}(\lambda)  &  = \kappa \alpha_{+}(\lambda)-\sigma
\lambda\alpha_{-}(\lambda)+2\sigma \hat{A}_{-}\left(  \lambda\right)  +\sigma
\hat{}A_{+}\left(  \lambda\right),  \label{eq:amSS}%
\end{align}
supplemented by the boundary conditions%
\begin{align}
\bar{\omega}_{+}\alpha_{+}(0)  &  =\bar{\nu},\\
\alpha_{\pm}(\infty)  &  =0.
\end{align}

Recalling that%
\begin{equation}
\alpha_{\pm}(\lambda)=-\frac{d}{d\lambda}\hat{A}_{\pm}\left(  \lambda\right),
\label{eq:AtoaSS}%
\end{equation}
we can cast the equations (\ref{eq:apSS}) and (\ref{eq:amSS}) solely in terms of $\hat{A}_{\pm}\left(  \lambda\right)$, viz.%
\begin{eqnarray}
-\bar{\omega}_{+}\frac{d^{2}}{d\lambda^{2}}\hat{A}_{+}(\lambda) =\left(
\kappa+\sigma\lambda\right)  \frac{d}{d\lambda}\hat{A}_{+}(\lambda)+\sigma \hat{A}_{+}\left(
\lambda\right)
=\frac{d}{d\lambda}\left\{\left(\kappa+\sigma\lambda\right)
\hat{A}_{+}(\lambda)\right\} \label{eq:ApSS}\\
\frac{d^{2}}{d\lambda^{2}}\hat{A}_{-}(\lambda)-\sigma\lambda\frac{d}{d\lambda
}\hat{A}_{-}(\lambda)-2\sigma A_{-}\left(  \lambda\right) =  -\kappa \frac{d}{d\lambda}%
\hat{A}_{+}(\lambda)+\sigma \hat{A}_{+}\left(  \lambda\right).  \label{eq:AmSS}%
\end{eqnarray}

To obtain relevant boundary conditions at $\lambda=0$, we recall that
$\hat{A}_{\pm}(0)=A^{(0)}_{\pm}.$ The moment equations in steady state yield for $n=0$%
\begin{align}
-\bar{\omega}_{+}\alpha_{+}(0)  &  =-\kappa A^{(0)}_{+}\label{eq:M0pSS}\\
\alpha_{-}(0)  &  =\kappa A^{(0)}_{+}+\sigma A^{(1)}_{-}+\sigma A^{(1)}_{+},
\label{eq:M0mSS}%
\end{align}
from which we find%
\begin{equation}
\hat{A}_{+}\left(  0\right)  =A^{(0)}_{+}=\frac{\bar{\nu}}{\kappa}.
\label{eq:Ap0SS}%
\end{equation}
For $n=1$ we have%
\begin{align}
-\bar{\omega}_{+}A^{(0)}_{+}  &  =-\kappa A^{(1)}_{+}-\frac{1}{2}\sigma A^{(2)}_{+}%
\label{eq:M1pSS}\\
A^{(0)}_{-}  &  = \kappa A^{(1)}_{+}+\frac{1}{2}\sigma A^{(2)}_{+},
\label{eq:M1mSS}%
\end{align}
from which we find%
\begin{equation}
\hat{A}_{-}\left(  0\right)  =A^{(0)}_{-}=\bar{\omega}_{+}A^{(0)}_{+}=\bar{\omega}%
_{+}\frac{\bar{\nu}}{\kappa}.
\label{eq:Am0}%
\end{equation}
At $\lambda\rightarrow\infty$ we simply have%
\begin{equation}
\hat{A}_{+}(\infty)=\hat{A}_{-}(\infty)=0.
\end{equation}

Equation (\ref{eq:ApSS}) can be integrated once to yield%
\begin{equation}
-\bar{\omega}_{+}\frac{d}{d\lambda}\hat{A}_{+}(\lambda)=\left(  \kappa+\sigma
\lambda\right)  \hat{A}_{+}(\lambda)+C_{0}. 
\label{eq:Ap1}%
\end{equation}
Using the result (\ref{eq:Ap0SS}) and the definition (\ref{eq:AtoaSS}), we find %
\begin{equation}
-\bar{\omega}_{+}\frac{d}{d\lambda}\hat{A}_{+}(0)= \kappa \hat{A}_{+}(0)+C_0 = \bar{\nu}+C_0= \bar{\omega}_{+}\alpha_{+}%
(0)=\bar{\nu},
\end{equation}
so that $C_{0}=0$.  The solution to (\ref{eq:ApSS}) is therefore given by%
\begin{equation}
\hat{A}_{+}(\lambda)=\frac{\bar{\nu}}{\kappa}e^{-\frac{1}{\bar{\omega}_{+}}\lambda\left(
\kappa+\frac{1}{2}\sigma\lambda\right)  }.
\label{eq;ApSS}%
\end{equation}

The solution Eq.\ (\ref{eq:AmSS}), an inhomogeneous ODE of degree 2, is more cumbersome to obtain. In principle, it can be solved using the variation-of-constants method. In practice, it turns out to be convenient to represent the solution as
\begin{equation}
\hat{A}_{-}(\lambda)=\hat{A}_{+}(\lambda)\chi\left(  \lambda\right)  ,
\label{eq:AMAp}%
\end{equation}
which divides out the exponentials in the inhomogeneous term. This leads to the following equation for
$\chi\left(\lambda\right)  $%
\begin{multline}
\left( \bar{\omega}_{+}\left(\kappa ^2+\kappa  \lambda  \sigma +\sigma  \bar{\omega}_{+} \right)\right)\\ +
 \left(-\lambda  \sigma  \bar{\omega}_{+}  (\kappa +\lambda  \sigma )-(\kappa +\lambda  \sigma )^2+2 \sigma  \bar{\omega}_{+} ^2+\sigma  \bar{\omega}_{+}\right)\chi(\lambda)\\+
 \left(\bar{\omega}_{+}  (2 \kappa +\lambda  \sigma  (\bar{\omega}_{+} +2))\right)\chi'(\lambda)\\
 -\bar{\omega}_{+}^2 \chi''(\lambda)=0
\end{multline}
With the aid of Mathematica \cite{Mathematica}, the solution to this equation
with boundary conditions $\chi(0)  =\bar{\omega}^{+}$ \ (cf. Eq.\ (\ref{eq:Am0})) and $\lim_{\lambda\rightarrow\infty}A^{+}(\lambda)\chi\left(  \lambda\right)=0$, is found to be%
 \begin{equation}
   \chi\left(\lambda\right)=\bar{\omega}_{+} -\frac{1}{2}\sqrt{\pi} \lambda  \sqrt{2 \sigma \bar{\omega}_{+} (\bar{\omega}_{+} +1)}  e^{\frac{(\kappa +\lambda  \sigma  (\bar{\omega}_{+} +1))^2}{2 \sigma  \bar{\omega}_{+}  (\bar{\omega}_{+} +1)}}
   \text{erfc}\left(\frac{\kappa +\lambda  \sigma  (\bar{\omega}_{+} +1)}{\sqrt{2 \sigma \bar{\omega}_{+} (\bar{\omega}_{+} +1)}}\right) 
\end{equation}

To obtain an equation for the steady-state pool size $\Lambda_{G},$ we start from the length conservation equation
\begin{equation}
\Lambda=\Lambda_{G}+\Lambda_{+}+\Lambda_{-}=\Lambda_{G}+A^{(1)}_{+}+A^{(1)}_{-}.%
\end{equation}
The moment equation Eq.\ (\ref{eq:M0mSS}) shows that%
\begin{equation}
A^{(1)}_{+}+A^{(1)}_{-}=\frac{1}{\sigma}\left( \alpha_{-}\left(  0\right)
-\kappa A^{(0)}_{+}\right) =-\frac{1}{\sigma}\left(\frac{d}{d\lambda}\hat{A}_{-}%
(0)+\bar{\nu}\right)  ,
\end{equation}
which can be readily evaluated using the explicit solution for $\hat{A}_{-}(\lambda)$. This leads to the following explicit self-consistency equation for the pool size%
\begin{align}
\Lambda &  =\Lambda_{G}+\frac{1}{2\sigma}\sqrt{\pi}\nu\left(\Lambda
_{G}\right)  \frac{e^{\Omega^{2}\left(  \Lambda_{G}\right)  }%
\operatorname{erfc}\left(  \Omega\left(  \Lambda_{G}\right)  \right)  }%
{\Omega\left(  \Lambda_{G}\right)  }\equiv\Lambda_{G}+\nu(\Lambda_G)\Phi\left(\Omega\left(\Lambda_{G}\right)\right)
,\label{eq:LamSS}\\
\Omega\left(\Lambda_{G}\right)   &  =\frac{\kappa}{\sqrt{2\sigma\omega_{+}\left(
\Lambda_{G}\right)  (\omega_{+}\left(  \Lambda_{G}\right)  +1)}},
\label{eq:OmegaSS}%
\end{align}
where we have replaced the `placeholders' $\bar{\omega}_{+}$ and $\bar{\nu}$ with their explicit dependency on the G-actin pool. Considering the case without severing, we get
\begin{equation}\label{eq:Psi0}
    \Phi_0(\Lambda_G)=\lim_{\sigma\rightarrow 0} \Psi\left(\Omega\left(\Lambda_{G}\right)\right)=\frac{1}{\kappa^2}\omega_{+}\left(\Lambda_{G}\right)\left(\omega_{+}\left(\Lambda_{G}\right)+1\right),
\end{equation}
with the latter result straightforwardly verified by solving the for the steady state in the actin dynamical model with $\sigma=0$ at the outset.

\section{Details of stochastic simulations}\label{app:simulations}
We perform stochastic simulations of dynamical actin filaments with a fixed time step $\Delta t$. At each time step, the size of the free G-actin pool $L_G$ retrieved. This sets the current value of all quantities that depend on this value, either directly such as the growth speed $v_{+}$ and the nucleation rate $r_n$, or indirectly because a parameter is modulated by the pool-dependent feedback mechanism, which depending on the specific scenario can be $r_n^\infty$, $v_b^\infty$, $r_c$ or $r_s$. The appropriate capping probability for the time step is then determined as $P_c = r_c \Delta t$. Subsequently, the state and current length of each filament are retrieved, the severing probability of the filament $P_s = r_s l \Delta t$ is determined, and a uniform random number $r \in (0,1)$ is drawn. When the filament is in the (non-capped) growing state, if $r < P_c$, the filament is switched to the (capped) shrinking state, else if $r \in [P_c,P_c+P_s)$ the filament is severed at a random position along its length, producing a shrinking filament of length $l_s<l$ and a growing one of length $l-l_s$. In all other cases, the filament grows by a length $\Delta l_+=v_+ \Delta t$. When the filament is in the shrinking case and $r < P_s$, it is severed, producing two shrinking filaments of lengths $l_s$ and $l-l_s$ respectively. Otherwise, it shrinks by a length $\Delta l_{-} = v_ {-}\Delta t$. In the latter case, whenever the resulting length falls below $0$, the filament is marked for removal. After all filaments have been updated, the number of newly nucleated growing actin filaments is determined by sampling a Poisson distribution with the probability of nucleation of a single filament given by $P_n = r_n \Delta t$ using the Knuth algorithm \cite{KnuthThe2}. The fixed baseline parameters used throughout our simulations are shown in Table \ref{tab:baseline_pars}. For the minimal scenario discussed in Section \ref{sec:results_minimal} we used an upper value for the nucleation rate of $r_n(1)=\SI{210}{\per\second}$. 

Table \ref{tab:modulated_pars} collects the parameters that are modulated in the four scenarios discussed in Section \ref{sec:results_full}. To ensure that in the most critical case, which is severing, where the probability of occurrence scales with the filament length, the single timestep event-probability remains well below $0.01$, we adopt a time step $\Delta t =$ \SI{0.01}{\second}. 

In the state-switching simulations, we increase or decrease the total amount of actin in the system by a percentage $\delta L$ of the original total length $L$. The change in length is then initially applied to the free pool only, so that $L_G \rightarrow L_G+\delta L \times 0.01L$, while $L_F$ is unchanged. The system then evolves freely to adapt to the new total actin length. After a time interval $\Delta \tau$ the change is undone and the system is restored to its original total actin length. The parameters used for the results in Fig.\ \ref{fig:full_modulation} are shown in Table \ref{tab:pool_manipulation}.

\begin{table}[t]
\centering
\begin{tabular}{lcll}
\hline\noalign{\smallskip}
Parameter & Symbol & Value & Source \\
\noalign{\smallskip}\hline\noalign{\smallskip}
Bare nucleation rate        & $r_{n}^\infty$ & \SI{70}{\per\second} & Assumed\\
Bare polymerization speed   & $v_{b}^\infty$ & \SI{15.6}{\micro\meter\per\second}  & \cite{Blanchoin2014}\\
Depolymerization speed      & $v_{p} $ & \SI{0.1}{\micro\meter\per\second} & \cite{Mogilner2002}\\
Capping rate & $r_{c}$      & \SI{3}{\per\second}& \cite{Pollard2000}\\
Severing rate & $r_{s}$     & \SI{0.005}{\per\micro\meter\per\second}  & \cite{Gurel2014}\\
Crossover length & $L_{\ast}$ & \SI{2000}{\micro\meter} & Assumed\\
Total actin length & $L$    & \SI{8000}{\micro\meter} & Assumed \\
\noalign{\smallskip}\hline\noalign{\smallskip}
Unbinding rate effectors& $k_u$ & \SI{0.25}{\per\second} & Assumed\\
Binding rate effectors & $k_b$ & \SI{0.001}{\per\micro\meter\per\second} & Assumed\\
Total number of effector species & $B$ & \num{1000} & Assumed\\
Crossover number of effectors in the dose-response curve & $B_*$ & \num{90} & Assumed\\
Hill coefficient dose-response curve  effector & $h$ & \num{20} & Assumed\\
\noalign{\smallskip}\hline
\end{tabular}
\caption{Baseline parameter values for the stochastic simulations. Above the divider: basic F-actin dynamical parameters. Below the divider: parameters of the effector mechanism.}\label{tab:baseline_pars} 
\end{table}
\begin{table}[t]
    \centering
    \begin{tabular}{lcccc}
    \hline\noalign{\smallskip}
      Case   &   $\Delta \tau_{\downarrow}$ & $\delta\Lambda_{\downarrow}$ &   $\Delta \tau_{\uparrow}$ & $\delta\Lambda_{\uparrow}$\\
      \noalign{\smallskip}\hline\noalign{\smallskip}
        Nucleation      & \SI{15}{\second} & \SI{20}{\percent}  & \SI{8}{\second}   & \SI{-20}{\percent}\\
        Polymerization  & \SI{25}{\second} & \SI{20}{\percent}  & \SI{20}{\second}  & \SI{-20}{\percent}\\
        Capping         & \SI{50}{\second} & \SI{35}{\percent}  & \SI{10}{\second}  & \SI{-20}{\percent} \\
        Severing        & \SI{20}{\second} & \SI{7.5}{\percent} & \SI{30}{\second}  & \SI{-10}{\percent}
    \end{tabular}
    \caption{Values for the duration ($\Delta\tau$) and magnitude ($\delta \lambda$) of the manipulations of the G-actin pool used in Section \ref{sec:results_full} to achieve the switching from the high-$\Lambda_F$ state to the low-$\Lambda_F$ state ($\downarrow$), or vice versa ($\uparrow$).}\label{tab:pool_manipulation}
\end{table}
\newpage
\bibliography{references}
\end{document}